\documentclass[pra,superscriptaddress,nofootinbib]{revtex4}

\usepackage{amsmath}
\usepackage{amsfonts}
\usepackage{amssymb}
\usepackage{graphicx}

\usepackage{hyperref}
\usepackage{booktabs}
\usepackage{hyperref}
\usepackage{cleveref}
\usepackage{color}
\usepackage{float}

	\newcommand{\ket}[1]{\left| #1 \right\rangle}
	\newcommand{\bra}[1]{\left\langle #1 \right|}

		\usepackage{color}

\begin{document}

\title{Routing a quantum state in a bio-inspired network}
\author{Elham Faraji}
\affiliation{School of Science and Technology, University of Camerino, I-62032 Camerino, Italy}
\affiliation{Aix Marseille Univ, Université de Toulon, CNRS, CPT, Marseille, France}
\affiliation{CNRS Centre de Physique Théorique UMR7332, 13288 Marseille, France}
\affiliation{INFN Sezione di Perugia, I-06123 Perugia, Italy}
\author{Alireza Nourmandipour}
\affiliation{Department of Physics, Sirjan University of Technology, 7813733385 Sirjan, Iran}
\author{Stefano Mancini}
\affiliation{School of Science and Technology, University of Camerino, I-62032 Camerino, Italy}
\affiliation{INFN Sezione di Perugia, I-06123 Perugia, Italy}
\author{Marco Pettini}
\affiliation{Aix Marseille Univ, Université de Toulon, CNRS, CPT, Marseille, France}
\affiliation{CNRS Centre de Physique Théorique UMR7332, 13288 Marseille, France}
\author{Roberto Franzosi}
\affiliation{DSFTA, University of Siena, Via Roma 56, 53100 Siena, Italy}
\affiliation{INFN Sezione di Perugia, I-06123 Perugia, Italy}
\affiliation{QSTAR and INO-CNR, largo Enrico Fermi 2, I-50125 Firenze, Italy}

\begin{abstract}
We consider a spin network resembling an $\alpha$-helix structure and study quantum information transfer over this bio-inspired network. The model we use is the Davydov model in its elementary version without a phononic environment. We investigate analytically and numerically the perfect state transfer (PST) in such a network which provides an upper bound on the probability of quantum states transfer from one node to another. We study PST for different boundary conditions on the network and show it is reachable between certain nodes and with suitable spin-spin couplings.
\end{abstract}

\date{\today }

\maketitle

\section{Introduction}
\label{INT}

In extensive networking quantum information processing, reliable communication represents one of the most prominent challenges to overcome  \cite{RevModPhys.86.1203}. From this perspective, the problem of high-fidelity quantum state/information transfer and quantum-network engineering, have been largely studied either theoretically \cite{PhysRevX.9.011006} or experimentally \cite{PRXQuantum.1.020317}.

In addition to being an essential building block for the progress of quantum-information technologies applications \cite{Gisin2007,Ladd2010,Pirandola2015}, such as quantum cryptography, quantum computation and teleportation, the investigation of optimal state transfer represents also a valuable tool for the investigation of the fundamental physics \cite{Hou2016,PhysRevLett.106.040403}.
A network of spins \cite{PhysRevLett.91.207901} interacting by an appropriate Hamiltonian, such as the $XY$ or the Heisenberg one \cite{PhysRevA.77.022315}, is a paradigm for the realization of long-distance communication devices. Here, the information encoded in the excitations of the network of spins can be shared between the parties, provided that one has quantum state transfer with high fidelity. In these cases, quantum information high fidelity transfer origins from suitable quantum interference effects induced from the network dynamics \cite{Casaccino2009}.

Nowadays, experimentalists are faced with the problem of synthesizing spin networks reliable to perform high-performance state transfer. For instance, in a possible scheme, qubits should be stored in quantum chips whose dimensions are comparable to one of the classical devices, and which are confined into hosting-equipments, at near absolute-zero temperature, to maintain quantum coherence.
Furthermore, the engineering of a chain of qubits is challenging. A possible setup, consider trapping the (two-level) atoms in a cavity. In this case, an external classical field confines the atoms in the minima of a profound optical-lattice potential, whereas the atoms interact via cavity-induced atom-atom interactions \cite{PhysRevA.105.052439}.

Recent advances in quantum science, have pioneered the way towards the description of biological realm phenomena in the framework of quantum mechanics. Indeed, quantum mechanics suitably describes properties of systems at nano-scale or lower, and many life processes, such as photosynthesis, vision, and respiration \cite{doi:10.1098/rsif.2018.0640}, take place at these scales. For instance, it has been proposed that the nuclear spins of phosphorus atoms could act as qubits in the brain which would enable it to function like a quantum computer \cite{FISHER2015593}. 
Also, new evidences suggest that quantum effects help plants to turn sunlight into fuel \cite{ENGEL2011222}. There are strong witnesses that the migratory birds have a sort of quantum compass which helps them to exploit the earth's magnetic fields for navigation \cite{Ritz2004}. Furthermore, examining the quantum mechanics' role in the matter and energy of living microorganisms could pave the way for our understanding of the origin of life.

Remarkably, most processes in which the quantum nature emerges, involve energy transfer to the living matter. Furthermore, there are considerable shreds of evidence that, the light-initiated reactions in living organisms, even at ambient conditions, display quantum-coherent behavior \cite{Olaya-Castro2012}. In the simplest scenario, enhancement in electron transport is driven by quantum-coherent processes where electronic excitations with a well-defined phase relationship at the molecular scale take place. In this regard, there are no better candidates than electrons for energy transfer, to understand the relevant quantum effects in living matter. Many works have been devoted to studying energy and information transmission efficiency in living organizations, such as in photosynthetic processes, in the framework of the quantum information processing \cite{Mohseni2008,PhysRevA.81.062346,Engel2007}. 
Several possible theoretical frameworks have been proposed to understand the nature of exciton propagation on a lattice \cite{Frohlich1950,HOLSTEIN1959325}, where the electron distribution belongs to this class of phenomena. In a seminal work, Davydov proposed that the mechanism for energy storage and transport in biomolecules could be considered soliton formation and propagation. In this scenario, the self-localization of quantum units of peptide vibrational energy might be the result of interactions with lattice phonons \cite{davydov1985solitons, SCOTT19921}. Especially, he suggested this mechanism in protein $\alpha$-helix which is composed of three channels of one-dimensional spring chain coupled with each other. 

Motivated by these considerations, in the present work we aim to investigate the quantum state transfer in a bio-inspired network of qubits. In particular, we consider an $\alpha$-helix structure. The latter looks like a two-dimensional spin network with the pair $(n,\alpha)$ denoting the $n$-th molecule along a chain and 
$\alpha$-th particular channel. In this model, the whole Hamiltonian is composed of three parts: the exciton energy operator, phonon energy operator, and exciton-phonon interaction operator. Usually, Davydov's phonon is a large acoustic polaron at low temperatures. Therefore, only the exciton part gives rise to a suitable candidate for the spin network. Given the fact that the propagation of quantum information over a network of spins is of great importance, we focus here only on the exciton part of Davydov's model. This in turn gives rise to four configurations based on the boundary conditions (open and closed) on both $n$ and $\alpha$. We illustrate in detail that only in one boundary condition analytical results for information transfer fidelity are obtainable. Anyway, we numerically show perfect state transfer also for other boundary conditions, but with closed boundaries for $n$ and $\alpha$.

The rest of the paper is organized as follows: In Sec. \ref{sec-intro} we present the model and discuss the transfer fidelity and its properties. Section \ref{closenalpha} is devoted to the special case of closed boundaries on both $n$ and $\alpha$, where an analytical investigation is possible. Secs. \ref{opennalpha},\ref{opennclosea} and \ref{closenopena} deal with the other possible boundary conditions by numeric calculation. Finally, we draw our conclusion in Sec. \ref{Conclusion}.

\section{The Model}\label{sec-intro}

\subsection{Davydov Model}
We consider a system with an $\alpha$-helix structure, which is the most stable configuration of the polypeptide \cite{Errington2006}. Due to the existence of three hydrogen bonds between peptide groups, the $\alpha$-helix usually looks like a 3-channel conformation each composed of a chain of springs coupled with each other. The Hamiltonian describing this system is usually referred to as Davydov's model.
Its bare form, without phonons reads  \cite{DAVYDOV1973559}
\begin{eqnarray}\label{main Hamiltonian}
\hat{H}_{\text{ex}}&=&\sum_{n=1}^{N}\sum_{\alpha=1}^{3}
\Big[E_{0}\hat{B}_{n,\alpha}^{\dag} \hat{B}_{n,\alpha}
+J\left(\hat{B}_{n,\alpha}^{\dag} \hat{B}_{n+1,\alpha}+\hat{B}_{n,\alpha}^{\dag} \hat{B}_{n-1,\alpha}\right)
+L\left(\hat{B}_{n,\alpha}^{\dag}\hat{B}_{n,\alpha+1}+ \hat{B}_{n,\alpha}^{\dag}\hat{B}_{n,\alpha-1}\right)\Big],
\end{eqnarray}
where $\hat{B}_{n,\alpha}$ and $\hat{B}^{\dag}_{n,\alpha}$ ($n=1,\ldots,N$, $\alpha=1,2,3$) are the spin-$\frac{1}{2}$ lowering and raising operators. The subscript $n$ labels the molecule along a macromolecule, while the subscript $\alpha$ specifies a particular channel. Furthermore, $E_0$ is the energy of exciton (amide-I vibration), $J$ is the nearest neighbour dipole-dipole coupling energy along a channel which is considered positive for an excited electron \cite{DAVYDOV19811} and $L$ is the nearest neighbour dipole-dipole coupling energy between channels. Thus, the pair $(n,\alpha)$ chooses an individual amino acid. The first term can be omitted in the interaction picture since it amounts to a conserved quantity.

This system can be interpreted in terms of a spin network. Let us consider a simple undirected graph (that is, without loops or parallel edges) $G=(V, E)$, in which the vertices (nodes) $V(G)=\{1,\ldots,3N\}$ are associated to the quantum spins. The edges $E(G)$ denote their allowed couplings.

The dimension ($2^{3N}\times 2^{3N}$) of the whole system's Hilbert space ${\cal H}\simeq (\mathbb{C}^2)^{\otimes 3N}$, makes pretty soon impracticable a numeric approach when $N$ is increased. Nevertheless, the above-mentioned conserved quantity, $\hat{Q}=\sum_{n=1}^{N}\sum_{\alpha=1}^{3} \hat{B}_{n,\alpha}^{\dag} \hat{B}_{n,\alpha}$, allows us to restrict our attention to the invariant single-excitation subspace ${\cal H}_1=span\{|n\rangle\}_{n=1}^{N}\bigotimes \{|\alpha\rangle\}_{\alpha=1}^{3}$. Here, the vector $|n,\alpha\rangle=|0\rangle_{1,1}|0\rangle_{1,2}\ldots|1\rangle_{n,\alpha}\ldots|0\rangle_{N,3}$ indicates the presence of excitation in the $n$-th molecule and $\alpha$-th channel. The $3N$ single-excitation states are conveniently labelled by an integer $n$ running from $0$ to $N-1$ within the three channels $\alpha=1,2,3$. In this new basis, the $3N\times 3N$ Hamiltonian (\ref{main Hamiltonian}) describing the system is represented by the matrix
\begin{equation}
\label{Hamilton}
\hat{H}_{\text{ex}} = 
\begin{pmatrix}
\textbf{C}_{0}^{0} & \textbf{C}_{1}^{0}& \textbf{C}_{2}^{0} & \cdots & \textbf{C}_{N-1}^{0} \\
\textbf{C}_{0}^{1}  & \textbf{C}_{1}^{1} & \textbf{C}_{2}^{1} & \cdots & \textbf{C}_{N-1}^{1} \\
\vdots   &    & \ddots &  \cdots  & \vdots  \\
\textbf{C}_{0}^{N-1}  & \textbf{C}_{1}^{N-1} & \textbf{C}_{2}^{N-1} & \cdots  & \textbf{C}_{N-1}^{N-1}  
\end{pmatrix},
\end{equation}
in which $\textbf{C}_{m}^{n}$ (with $m,n=0,\cdots N-1$) are $3\times 3$ matrices which depend on the boundary conditions. In the next sections, we consider the possible boundary conditions in details. 

\subsection{Information Transfer Fidelity}
\label{MTF}

The Hamiltonian (\ref{Hamilton}) can be written as its spectral decomposition as follow
\begin{eqnarray}\label{decompositionGen}
\hat{H}_{\text{ex}}=\sum_{n=0}^{N-1}\sum_{\alpha=1}^{3}\lambda_{n}^{\alpha}
\begin{matrix}
\hat{\prod}_{n}^{\alpha}
\end{matrix},
\end{eqnarray}
in which $\lambda_{n}^{\alpha}$ are the eigenvalues of the Hamiltonian with their corresponding eigenvector $\ket{W_{n}^{\alpha}}$ so that we have $\hat{\prod}_{n}^{\alpha}=\ket{W_{n}^{\alpha}}\bra{W_{n}^{\alpha}}$. Suppose the single excitation initially resides in the $i$th qubit at channel $p$, i.e., $\ket{\psi(0)}=\ket{i,p}$ with $i\in\{0,1,\cdots,N-1\}$ and $p\in\{1,2,3\}$. Since $[\hat{Q},\hat{H}_{ex}]=0$, only the transitions of the form $\ket{i,p}\leftrightarrow\ket{j,q}$ are allowed. Therefore, we define the quantum transition probability from state $\ket{i,p}$ into $\ket{j,q}$ as follows 
\begin{equation}
\label{transitionpro}
p_{t}([i,p],[j,q])=|\langle i,p|e^{-i\hat{H}_{\text{ex}}t}|j,q\rangle|^2.
\end{equation}
Using the spectral decomposition (\ref{decompositionGen}), the fidelity between the evolved input state and the desired output is written as 
\begin{equation}
\label{decatt}
p_{t}([i,p],[j,q])=\left|\sum_{n=0}^{{N}-1}\sum_{\alpha=1}^{3}\langle i,p|
\begin{matrix}
\prod_{n}^{\alpha}
\end{matrix}
|j,q\rangle e^{-i\lambda_{n}^{\alpha}t}\right|^2.
\end{equation}
An upper bound for the transition probability is derived by
assuming that there exists $t$ such that the phase factor $e^{-i\lambda_{n}^{\alpha}t}$ can be written as 
\begin{eqnarray}\label{signatt}
e^{-i\lambda_{n}^{\alpha}t}=s_{n}^{\alpha}([i,p],[j,q])e^{i\phi},\;\;\;\;\forall n=0,...,N-1,\;\;\;\; \text{and}\;\;\;\; \alpha=1,2,3,
\end{eqnarray}
where either $s_{n}^{\alpha}([i,p],[j,q]):= \text{Sgn}(\langle i,p|\prod_{n}^{\alpha}|j,q \rangle)\in \{-1,+1\}$ as a sign factor or   $s_{n}^{\alpha}([i,p],[j,q])=0$ whenever $\langle i,p|\prod_{n}^{\alpha}|j,q \rangle=0$  and $\phi$ is an arbitrary global phase which must be the same for all $n$'s and $\alpha$'s. In this case, since the global phase $\phi$ can be absorbed in the absolute value in (\ref{decatt}), the upper bound for the transition probability is obtained as follow \cite{Jonckheere2015}
\begin{equation}\label{ITF}
p_{\text{max}}([i,p],[j,q])=\Big(\sum_{n=0}^{N-1}\sum_{\alpha=1}^{3}|\langle i,p|
\begin{matrix}
\prod_{n}^{\alpha}
\end{matrix}
|j,q\rangle|\Big)^2.
\end{equation}
The maximum amplitude for the transition probability $p_{\text{max}}([i,p],[j,q])$ is $1$ and this case is referred to as \textit{Perfect State Transfer} (PST), see e.g. \cite{Franzosi_2014,6044395}. Furthermore, due to the exponential nature of the phase factor $e^{-i\lambda_{n}^{\alpha}t}$ there must exist a sequence of time samples $t_n$ for which PST occurs. 

\begin{figure}[H]
              \includegraphics[width=0.23\columnwidth]{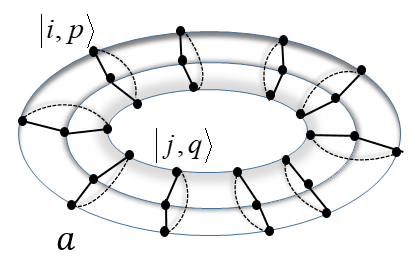}
             \includegraphics[width=0.23\columnwidth]{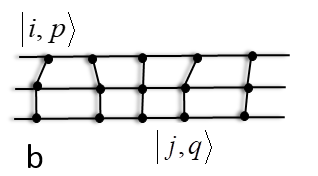}
                 \includegraphics[width=0.23\columnwidth]{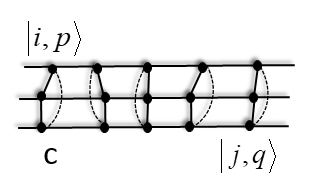}
                 \includegraphics[width=0.23\columnwidth]{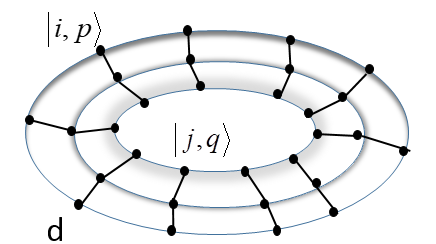}
                  \caption{The boundary conditions: a) Closed  on both $N$ and $\alpha$, b) Open on both $N$ and $\alpha$, c) Open on $N$ and closed on $\alpha$, and d) Closed on $N$ and open on $\alpha$.}
                   \label{diffboundaries}
                \end{figure}

According to Eq. (\ref{signatt}), there exists eigenspaces with $s_{n}^{\alpha}([i,p],[j,q])=0$ which do not give any contribution in the sum (\ref{ITF}), thus they can be removed from the summation. We refer to them as \textit{dark-state} subspaces. Then we can limit ourselves to the $K'\subseteq\{0,1,...,N-1\}$ of indices $n$ for which $s_{n}^{\alpha}([i,p],[j,q])\ne 0$. For simplicity let us use $s_n^{\alpha}$ instead of $s_n^{\alpha}([i,p],[j,q])$. Then we point out that for all members of set $K'$ we have $s_n^{\alpha}=\pm 1$ and $\exp\left[ -i\frac{\pi}{2}(s_n^{\alpha}-1)\right]=\pm 1 $. Therefore, we may write $s_n^{\alpha}=\exp\left[ -i\pi\left( 2k_n^{\alpha}+\frac{1}{2}(s_n^{\alpha}-1)\right) \right]$ where $k_n^{\alpha}\in \mathbb{Z}$ are arbitrary integers. Using this relation and Eq. (\ref{signatt}) we may obtain the following relation
\begin{equation}
\lambda_n^{\alpha}t=2\pi k_n^{\alpha}+\frac{\pi}{2}(s_n^{\alpha}-1)-\phi ),\;\;\;\forall n \in K'\;\;\text{and}\;\; \alpha=1,2,3.
\label{lambda}
\end{equation}
In a pairwise manner with $[n,\alpha] \ne [m,\beta]$, the above attainability condition becomes  
\begin{eqnarray}\label{attainability1}
(\lambda_{n}^{\alpha}-\lambda_{m}^{\beta})t=2\pi (k_{n}^{\alpha}-k_{m}^{\beta})+\frac{\pi}{2}(s_{n}^{\alpha}-s_{m}^{\beta}),\;\;\;\forall n,m \in K'\;\;\text{and}\;\; \alpha,\beta=1,2,3,
\end{eqnarray}
which is more useful as it is independent of the arbitrary phase $\phi$. In the above relation $k_{n}^{\alpha},k_{m}^{\beta}\in \mathbb{Z}$ are arbitrary integers. Noting $s_{n}^{\alpha}$ and $s_{m}^{\alpha}=\pm 1$, we can write the attainability constraint (\ref{attainability1}) as
\begin{equation}\label{attainability2}
\begin{aligned}
(\lambda_{n}^{\alpha}-\lambda_{m}^{\beta})t&= 2\pi(k_{n}^{\alpha}-k_{m}^{\beta}),\;\;\;\;\;\;\;\;\;\;\;\;\;\;\;\text{if}\;\;s_{n}^{\alpha}=s_{m}^{\beta},\\
(\lambda_{n}^{\alpha}-\lambda_{m}^{\beta})t&= 2\pi (k_{n}^{\alpha}-k_{m}^{\beta})+\pi,\;\;\;\;\;\;\;\;\text{if}\;\;s_{n}^{\alpha}=-s_{m}^{\beta}=1,\\
(\lambda_{n}^{\alpha}-\lambda_{m}^{\beta})t&= 2\pi (k_{n}^{\alpha}-k_{m}^{\beta})-\pi,\;\;\;\;\;\;\;\;\text{if}\;\;s_{n}^{\alpha}=-s_{m}^{\beta}=-1.
\end{aligned}
\end{equation}
Note that from above equations, we look for PST between the particular input and output states  $([i,p],[j,q])$ (through the values of $s_{n}^{\alpha}$ and $s_{m}^{\beta}$), while the pair $([n,\alpha],[m,\beta])$ indicates two arbitrary nodes in the network (according to the summation in Eq. (\ref{decatt})). 
The equations (\ref{attainability2}) imply that if (\ref{attainability1}) holds for the couples $([n,\alpha],[m,\beta])$ and $([m,\beta],[r,\eta])$ then (\ref{attainability1}) holds also for $([n,\alpha],[r,\eta])$, with $r\in K'$ and $\alpha,\beta,\eta=1,2,3$. Furthermore, the set of equations are redundant, thus, in order to consider just linear independent equations, we restrict ourselves to an appropriate subset of equations (\ref{attainability2}) with $([n,\alpha],[n,\beta])$ where $\alpha\neq \beta$ and $([n,\alpha],[n+1,\beta])$ for $n \in K'$ and $\alpha,\beta=1,2,3$. Moreover, note that the conditions (\ref{attainability2}) are sufficient but not necessary for PST because they guarantee the attainability of the upper bound $p_{max}$ in (\ref{ITF}), but it is not guaranteed that this is equal to $1$ (i.e. that it gives PST).\\
We are in the position to apply the above-described tools, for studying the information transmission in different configurations of the spin network, determined by the possible boundary conditions. 
Figure  \ref{diffboundaries} illustrates the topologies of the different cases investigated. The case of closed boundary conditions either along the chain and the channels is sketched in Fig. \ref{diffboundaries} $a$, the configuration with open boundary conditions along the chain and the channels is illustrated in Fig. \ref{diffboundaries} $b$, the case of open (closed) boundary conditions along the chain and closed (open) conditions for the channels, is described in Fig. \ref{diffboundaries} $c$ (Fig. \ref{diffboundaries} $d$).

\section{Closed boundary on both $N$ and $\alpha$}\label{closenalpha}

In this section, we consider the case of closed boundary conditions on both $n$ and $\alpha$ (see figure \ref{diffboundaries} $(a)$). Restricted to the single excitation subspace, the Hamiltonian of the system becomes a block circulant matrix with circulant blocks (which are also symmetric blocks). In fact, $\hat{H}_{\text{ex}}$ is a $3N \times 3N$ block circulant matrix with $3 \times 3$  circulant matrices $ \textbf{C}_{0}^{0}, \textbf{C}_{1}^{0},..., \textbf{C}_{N-1}^0$, where
\begin{eqnarray}\label{matrixH}
\textbf{C}_{0}^0=\left(
      \begin{array}{ccc}
       0 & L& L \\
        L & 0& L \\
        L & L& 0\\
       \end{array}
     \right);\;\;\;\;\;\textbf{C}_{1}^0=\textbf{C}_{N-1}^0=\left(
      \begin{array}{ccc}
       J & 0& 0 \\
        0 & J& 0 \\
        0 & 0& J\\
       \end{array}
     \right)
\end{eqnarray}
and $\textbf{C}_{2}^0=\textbf{C}_{3}^0=...=\textbf{C}_{N-2}^0=0$. Then the Hamiltonian $\hat{H}_{\text{ex}}$ has the matrix form
\begin{equation}
\label{BothClose}
\hat{H}_{\text{ex}} = 
\begin{pmatrix}
\textbf{C}_{0}^0 & \textbf{C}_{1}^0& 0 & \cdots & \textbf{C}_{N-1}^0 \\
\textbf{C}_{N-1}^0  & \textbf{C}_{0}^0 & \textbf{C}_{1}^0 &  & 0 \\
0 & \textbf{C}_{N-1}^0 & \ddots & \ddots & \vdots \\
\vdots   &    & \ddots &    & \textbf{C}_{1}^0  \\
\textbf{C}_{1}^0  & 0 & \cdots & \textbf{C}_{N-1}^0  & \textbf{C}_{0}^0  
\end{pmatrix}.
\end{equation}
The eigenvalues of $\hat{H}_{\text{ex}}$  are \cite{tee2007eigenvectors}
\begin{eqnarray}\label{eigenvales}
\lambda_{n}^{\alpha}=2J \cos{({2\pi n\over N})}-2e^{i\alpha\pi}L\cos{({\pi \over 3}(\alpha-1))},
\end{eqnarray}
in which $\lambda_n^\alpha$ is the $\alpha$th diagonal element of the $n$th $3\times 3$ block for $n=0,1,2,...,N-1$ and  $\alpha=1,2,3$. Moreover, the normalized eigenvectors of $\hat{H}_{\text{ex}}$ are found to be \cite{tee2007eigenvectors}
\begin{eqnarray}\label{eigenvectorsN}
W_{n}^{\alpha}=\frac{1}{\sqrt{3N}}\Big(\rho_{n}^{-1}V_{\alpha},V_{\alpha},\rho_{n}V_{\alpha},\rho_{n}^{2}V_{\alpha},...,\rho_{n}^{N-2}V_{\alpha}\Big)^{T}
\end{eqnarray}
where $\rho_{n}=e^{i2\pi n/N}$ is the $N$-th Root of Unity for $n=0,1,...,N-1$ and $V_{\alpha}$ is defined as
\begin{eqnarray}
{V}_{1}=\Big(1,1,1\Big)^{T},\;\;\;\;\; {V}_{2}=\Big(e^{-\frac{2i\pi}{3}},1,e^{\frac{2i\pi}{3}}\Big)^{T},\;\;\;\;\; {V}_{3}=\Big(e^{\frac{2i\pi}{3}},1,e^{-\frac{2i\pi}{3}}\Big)^{T}.
\end{eqnarray}
    
\subsection{Eigendecomposition of the Hamiltonian}\label{EigendecompositionH}
We first point out that according to (\ref{eigenvales}), the eigenvalues corresponding to $\alpha=2$ and $\alpha=3$ are equal, i.e., $\lambda_n^2=\lambda_n^3$. This allows us to discuss the problem with the eigendecomposiotion of the Hamiltonian for only two cases corresponding to $\alpha=1$ and $\alpha=2$ (or 3). \\
We begin by considering the case $\alpha=1$:
\begin{itemize}
\item
If $N$ is even, but not divisible by $4$, we have $\lambda_{n}^{1}=\lambda_{N-n}^{1}=-\lambda_{N/2-n}^{1}=-\lambda_{N/2+n}^{1}\ne 0$,
 which means there are $\frac{N}{2}-1$ distinct pairs of double eigenvalues and two single eigenvalues $\pm 2J+2L$ for $n=0$ and $n=\frac{N}{2}$ giving totally $\tilde{N}=\frac{N}{2}+1$ pairwise distinct eigenvalues:
\begin{eqnarray}\label{even1}
\{-2J+2L,\lambda_{n}^{1},2J+2L:n=1,2,...,\frac{N}{2}-1\}.
\end{eqnarray}
In this case, if $N$ is divisible by $4$, there is also a double eigenvalues at $0$ for $n=\frac{N}{4}$ and $\frac{3N}{4}$ giving totally  $\tilde{N}=\frac{N}{2}+2$ pairwise distinct eigenvalues.
\item
If $N$ is odd, $\lambda_{n}^{1}=\lambda_{N-n}^{1}\ne 0$ and there are $(N-1)/2$ distinct pairs of double eigenvalues and a single eigenvalue $2J+2L$ at $n=0$ which is totally $\tilde{N}=(N+1)/2$ distinct eigenvalues:
\begin{eqnarray}\label{odd1}
\{\lambda_{n}^{1},2J+2L:n=1,2,...,\frac{N-1}{2}\}.
\end{eqnarray}
\end{itemize}

On the other hand, considering the cases $\alpha=2$ (or $\alpha=3$), we have:
\begin{itemize}
\item
If $N$ is even, but not divisible by $4$, we have $\lambda_{n}^{\alpha}=\lambda_{N-n}^{\alpha}=-\lambda_{N/2-n}^{\alpha}=-\lambda_{N/2+n}^{\alpha}\ne 0$ and so $\frac{N}{2}-1$ distinct pairs of the double eigenvalues and two single eigenvalues $\pm 2J-L$ giving totally $\tilde{N}=\frac{N}{2}+1$ pairwise distinct eigenvalues:
\begin{eqnarray}\label{even2}
\{-2J-L,\lambda_{n}^{\alpha},2J-L:\alpha=2\;\;\; (\text{or}\;\;\; 3)\;\;\; \text{and}\;\;\; n=1,2,...,\frac{N}{2}-1\}.
\end{eqnarray}
If $N$ is divisible by $4$, a double eigenvalue at $0$ for $n=\frac{N}{4}$ and $\frac{3N}{4}$ giving totally  $\tilde{N}=\frac{N}{2}+2$ pairwise distinct eigenvalues.
\item
If $N$ is odd, $\lambda_{n}^{\alpha}=\lambda_{N-n}^{\alpha}\ne 0$, we have $(N-1)/2$ distinct pairs of double eigenvalues and a single eigenvalue $2J-L$ at $n=0$ which is totally $\tilde{N}=(N+1)/2$ distinct eigenvalues:
\begin{eqnarray}\label{odd2}
\{\lambda_{n}^{\alpha},2J-L:\alpha=2\;\;\; (\text{or}\;\;\; 3)\;\;\; \text{and}\;\;\; n=1,2,...,\frac{N-1}{2}\}
\end{eqnarray}
\end{itemize}
Now we are in a position to construct the eigendecomposition of the Hamiltonian. To this end, we first define for the both even and odd $N$, the eigenprojection on the corresponding double eigenvalue $\lambda_{n}^{\alpha}=\lambda_{N-n}^{\alpha}$ to be as $\prod_{n}^{\alpha}:=|W_{n}^{\alpha}\rangle \langle W_{n}^{\alpha}|+| W_{N-n}^{\alpha}\rangle \langle W_{N-n}^{\alpha}|$. Also, we consider the eigenprojection of the eigenvalue at $n=0$ as $\prod_{0}^{\alpha}:=| W_{0}^{\alpha}\rangle \langle W_{0}^{\alpha}|$. Finally, if $N$ is even, the eigenprojection of the eigenvalue at $n=N/2$ is $\prod_{N/2}^{\alpha}:=| W_{N/2}^{\alpha}\rangle \langle W_{N/2}^{\alpha}|$ and further, if $N$ is divisible by $4$, then we define also $\prod_{N/4}^{\alpha}:=| W_{N/4}^{\alpha}\rangle \langle W_{N/4}^{\alpha}|+| W_{3N/4}^{\alpha}\rangle \langle W_{3N/4}^{\alpha}|$ for the corresponding eigenvalues at $n=N/4$ and $n=3N/4$. With this notation, the spectral decomposition (\ref{decompositionGen}) for the Hamiltonian (\ref{BothClose}) will be
\begin{eqnarray}\label{decomposition}
\hat{H}_\text{ex}=\sum_{n=0}^{\tilde{N}-1}\sum_{\alpha=1}^{3}\lambda_{n}^{\alpha}
\begin{matrix}
\prod_{n}^{\alpha}.
\end{matrix}
\end{eqnarray}
Then as mentioned above in (\ref{MTF}), to avoid having redundant equations of (\ref{attainability2}) we limit ourselves only to the linear independent equations for $([n,1],[n,2])$ where $n\in K'=\{0,1,...,\tilde{N}-1\}$ and $([n,\alpha],[n+1,\beta])$ where $n \in K'=\{0,1,...,\tilde{N}-2\}$ and $\alpha,\beta=1,2 \;(\text{or}\;\; 3).$

\subsection{Dark states}
In order to find the dark states of our $3N$-qubit system, we should study when $s_{n}^{\alpha}([i,p],[j,q])= 0$.  According to the subsection (\ref{EigendecompositionH}), we point out that for $n=0$, we have
\begin{equation}
\begin{aligned}
\langle i,p|\Pi_{0}^{1}|j,q \rangle &=1/3N\ne 0, \\
\langle i,p|\Pi_{0}^{2}|j,q \rangle&=e^{2i(p-q)\pi /3}/3N \ne 0, \\
\langle i,p|\Pi_{0}^{3}|j,q \rangle&=e^{2i(q-p)\pi /3}/3N \ne 0.
\end{aligned}
\end{equation}
This means there is no dark state with respect to $n=0$ (with multiplicity of 1) for any value of $\alpha$. Also, for an even $N$, there is the eigenvalue with multiplicity of 1 with respect to $n=N/2$. It is straightforward to illustrate the following relation for the case $n=N/2$:
\begin{equation}
\begin{aligned}
\langle i,p|\Pi_{N/2}^{1}|j,q \rangle &=(-1)^{(i-j)}/3N \ne 0, \\
\langle i,p|\Pi_{N/2}^{2}|j,q \rangle&=(-1)^{(i-j)}e^{2i(p-q)\pi /3}/3N \ne 0, \\
\langle i,p|\Pi_{N/2}^{3}|j,q \rangle&=(-1)^{(i-j)}e^{2i(q-p)\pi /3}/3N \ne 0.
\end{aligned}
\end{equation}
Therefore, there is no dark state associated with $n=N/2$ for an even $N$. Furthermore, for the eigenvalues with multiplicity $2$ (i.e., $\lambda_n^{\alpha}=\lambda_{N-n}^{\alpha} $ with $\alpha=1,2,3$) we have
\begin{equation}
\label{NEven}
\begin{aligned}
\langle i,p|\Pi_{n}^{1}+\Pi_{N-n}^{1}|j,q \rangle&=2\cos\left(2\pi n(j-i)/N\right)/3N,\\
\langle i,p|\Pi_{n}^{2}+\Pi_{N-n}^{2}|j,q \rangle&=2\cos \left(2\pi n(j-i)/N\right)e^{2i(p-q)\pi /3}/3N,\\
\langle i,p|\Pi_{n}^{3}+\Pi_{N-n}^{3}|j,q \rangle&=2\cos\left(2\pi n(j-i)/N\right)e^{2i(q-p)\pi /3}/3N,
\end{aligned}
\end{equation}
for $n=1,...,\left \lceil{(N-3)/2}\right \rceil$. According to the relations (\ref{NEven}), it is obvious that for all values of $\alpha$, there are dark states if $\cos \left(2\pi n(j-i)/N\right)=0$, i.e. when $4n(j-i)/N$ is an odd integer. \color{black} Thus, we can say that the divisibility of $N$ by $4$ is a necessary but not sufficient condition to have dark states due to the term $(j-i)$ relevant to the input and output states.

Our calculations shows that $p_{max}([i,p],[j,q])$ in (\ref{ITF}) is always $1$ ( independent on input and output qubits) due to the fact that the analytical eigenvectors of the Hamiltonian (\ref{BothClose}) are the Kronecker product of two Fourier matrices (\ref{eigenvectorsN}) where each element of the eigenvectors has the same structure as $e^{i\phi}/\sqrt{3N}$.

In what follows we use the rescaled time $\tau=Lt$ and the dimensionless parameter $\gamma=J/L$ in our all analytical and numerical computations.

\begin{figure}[H]
 \centerline{\includegraphics[width=0.45\columnwidth]{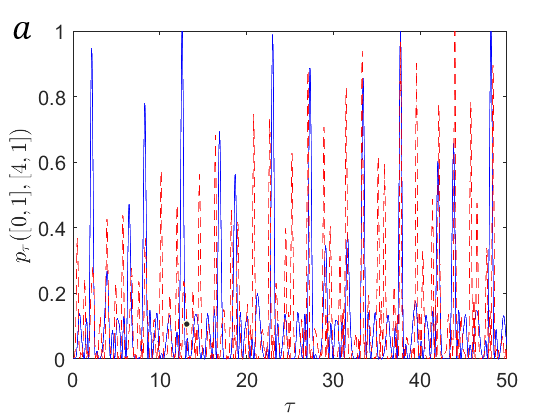}}
  \includegraphics[width=0.45\columnwidth]{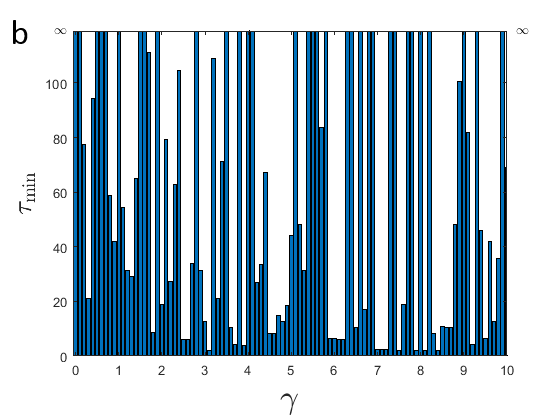}
   \includegraphics[width=0.45\columnwidth]{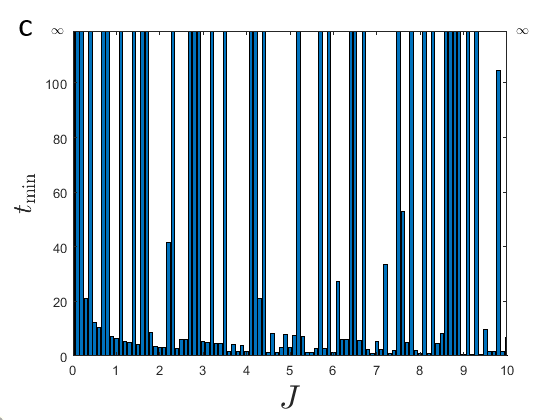}
   \caption{The transmission probability $p_{\tau}([0,1],[4,1])$ of a spin network with $N=8$ nodes and closed boundary on both $N$ and $\alpha$; a) $p_{\tau}$ vs $\tau$ for $\gamma=3$ (solid blue line) and $\gamma=5$ (red dashed line), b) the minimum value $\tau_{min}$ versus $\gamma$ at which $p_{\tau}=1$ for the finite values of $\gamma$, and c) the minimum value $t_{min}$ versus $J$ at which $p_{t}=1$ for $\gamma\rightarrow \infty$, i.e. $L\rightarrow 0$.}
\label{N8}
\end{figure}

\textbf{Example:} Now we aim to analytically explore the attainability conditions (\ref{attainability2}) of $p_{\tau}([i,p],[j,q])$ for $N=8$ qubits with the channels $\alpha=1,2 \ \text{and} \ 3$. According to the subsection (\ref{EigendecompositionH}), there are $6$ pairwise distinct eigenvalues, i.e., $\tilde{N}=6$. Choosing the suitable subset  $\mathcal{S}\subseteq K'\times K'$ where $K'=\{0,1,...,5\}$, the attainability condition (\ref{attainability1}) becomes 
  \begin{eqnarray}\label{attainabilitynm}
  (\lambda_{n}^{\alpha}- \lambda_{m}^{\beta})\tau&=&\tau\Big(-4\gamma\sin(\frac{\pi}{N}(n+m))\sin(\frac{\pi}{N}(n-m))\nonumber\\
  &-& 2\left(e^{i\alpha\pi}\cos(\frac{\pi}{3}(\alpha-1))-e^{i\beta\pi}\cos(\frac{\pi}{3}(\beta-1))\right)\Big);\;\forall n,m\in K'=\{0,1,...,5\};\;\text{and}\;\;\alpha,\beta\in\{1,2\}.
  \end{eqnarray}
Now in order to have linear independent equations, we consider several specific cases:
 \begin{itemize}
 \item  We first restrict ourselves to the subset $([n,\alpha], [n+1,\beta])$ where $n \in K'=\{0,1,...,\tilde{N}-2\}$ and $\alpha=\beta\in\{1,2\}$. From (\ref{attainabilitynm}) we obtain
  \begin{eqnarray}\label{example1condition1}
  (\lambda_{n}^{\alpha}- \lambda_{n+1}^{\alpha})\tau=-4\gamma\sin(\frac{\pi}{N}(2n+1))\sin(\frac{\pi}{N})\tau,\;\;\;\;\forall& n\in K'=\{0,1,2,3,4\};\;\text{and}\;\;\alpha\in\{1,2\}.
  \end{eqnarray} 
  Then from the attainability condition (\ref{attainability2}) we derive
  \begin{equation} 
  \label{example1condition11}
  \begin{aligned}
 (\lambda_{0}^{\alpha}- \lambda_{1}^{\alpha})\tau&=(2-\sqrt{2})\gamma \tau=2\pi(k_{0}^{\alpha}-k_{1}^{\alpha}),\;\;\;\;\;s_{0}^{\alpha}=s_{1}^{\alpha},\\
 (\lambda_{0}^{\alpha}- \lambda_{1}^{\alpha})\tau&=(2-\sqrt{2})\gamma \tau=2\pi(k_{0}^{\alpha}-k_{1}^{\alpha})+\pi,\;\;\;\;\;s_{0}^{\alpha}=-s_{1}^{\alpha}=1,\\
  (\lambda_{0}^{\alpha}- \lambda_{1}^{\alpha})\tau&=(2-\sqrt{2})\gamma \tau=2\pi(k_{0}^{\alpha}-k_{1}^{\alpha})-\pi,\;\;\;\;\;s_{0}^{\alpha}=-s_{1}^{\alpha}=-1,\\
   \end{aligned}
   \end{equation}
   and also
   \begin{equation} 
   \label{example1condition12}
   \begin{aligned}
     (\lambda_{1}^{\alpha}- \lambda_{2}^{\alpha})\tau&=\sqrt{2}\gamma \tau=2\pi(k_{1}^{\alpha}-k_{2}^{\alpha}),\;\;\;\;\;s_{1}^{\alpha}=s_{2}^{\alpha},\\
(\lambda_{1}^{\alpha}- \lambda_{2}^{\alpha})\tau&=\sqrt{2}\gamma \tau=2\pi(k_{1}^{\alpha}-k_{2}^{\alpha})+\pi,\;\;\;\;\;s_{1}^{\alpha}=-s_{2}^{\alpha}=1,\\
(\lambda_{1}^{\alpha}- \lambda_{2}^{\alpha})\tau&= \sqrt{2}\gamma \tau=2\pi(k_{1}^{\alpha}-k_{2}^{\alpha})-\pi,\;\;\;\;\;s_{1}^{\alpha}=-s_{2}^{\alpha}=-1,\\
  \end{aligned}
  \end{equation}
 Besides, $(\lambda_{0}^{\alpha}- \lambda_{1}^{\alpha})=(\lambda_{3}^{\alpha}- \lambda_{4}^{\alpha})=(\lambda_{4}^{\alpha}- \lambda_{5}^{\alpha})$ and $(\lambda_{1}^{\alpha}- \lambda_{2}^{\alpha})=(\lambda_{2}^{\alpha}- \lambda_{3}^{\alpha})$.

\color{black}
 \item  Secondly, we consider the restricted subset $([n,\alpha], [n+1,\beta])$ where $n \in K'=\{0,1,...,\tilde{N}-2\}$ while $\alpha\ne\beta\in\{1,2\}$. Then, from
    equation (\ref{attainabilitynm}) we have
     \begin{eqnarray} 
      (\lambda_{n}^{\alpha}- \lambda_{n+1}^{\beta})\tau&=&\tau \Big(-4\gamma\sin(\frac{\pi}{N}(2n+1))\sin(\frac{\pi}{N})\nonumber\\
      &-& 2\left(e^{i\alpha\pi}\cos(\frac{\pi}{3}(\alpha-1))-e^{i\beta\pi}\cos(\frac{\pi}{3}(\beta-1))\right)\Big)\nonumber\\
      &\forall& n\in K'=\{0,1,2,3,4\};\;\text{and}\;\;\alpha\ne\beta\in\{1,2\}.
      \end{eqnarray} 
      Then the condition (\ref{attainability2}) implies
   \begin{equation} \label{attainabiliy51}
   \begin{aligned}
  (\lambda_{0}^{\alpha}- \lambda_{1}^{\beta})\tau&=((2-\sqrt{2})\gamma+3)\tau=2\pi(k_{0}^{\alpha}-k_{1}^{\beta}),\;\;\;\;\;s_{0}^{\alpha}=s_{1}^{\beta},\\
  (\lambda_{0}^{\alpha}- \lambda_{1}^{\beta})\tau&=((2-\sqrt{2})\gamma+3)\tau=2\pi(k_{0}^{\alpha}-k_{1}^{\beta})+\pi,\;\;\;\;\;s_{0}^{\alpha}=-s_{1}^{\beta}=1,\\
   (\lambda_{0}^{\alpha}- \lambda_{1}^{\beta})\tau&=((2-\sqrt{2})\gamma+3)\tau=2\pi(k_{0}^{\alpha}-k_{1}^{\beta})-\pi,\;\;\;\;\;s_{0}^{\alpha}=-s_{1}^{\beta}=-1,\\
    \end{aligned}
    \end{equation}
    and also
       \begin{equation} 
       \begin{aligned}\label{attainabiliy52}
    (\lambda_{1}^{\alpha}- \lambda_{2}^{\beta})\tau&=(\sqrt{2}\gamma+3)\tau=2\pi(k_{1}^{\alpha}-k_{2}^{\beta}),\;\;\;\;\;s_{1}^{\alpha}=s_{2}^{\beta},\\
    (\lambda_{1}^{\alpha}- \lambda_{2}^{\beta})\tau&=(\sqrt{2}\gamma+3)\tau=2\pi(k_{1}^{\alpha}-k_{2}^{\beta})+\pi,\;\;\;\;\;s_{1}^{\alpha}=-s_{2}^{\beta}=1,\\
    (\lambda_{1}^{\alpha}- \lambda_{2}^{\beta})\tau&= (\sqrt{2}\gamma+3)\tau=2\pi(k_{1}^{\alpha}-k_{2}^{\beta})-\pi,\;\;\;\;\;s_{1}^{\alpha}=-s_{2}^{\beta}=-1.\\
      \end{aligned}
      \end{equation}
       Moreover, $(\lambda_{0}^{\alpha}- \lambda_{1}^{\beta})=(\lambda_{3}^{\alpha}- \lambda_{4}^{\beta})=(\lambda_{4}^{\alpha}- \lambda_{5}^{\beta})$ and $(\lambda_{1}^{\alpha}- \lambda_{2}^{\beta})=(\lambda_{2}^{\alpha}- \lambda_{3}^{\beta})$.\\

       \color{black}
 \item Finally, We choose the subset $([n,\alpha], [n,\alpha+1])$ where $n\in K' =\{0,1,...,\tilde{N}-1\}$ and $\alpha= 1$ to find other linear
 independent equations of (\ref{attainabilitynm}) given by
    \begin{eqnarray} 
    (\lambda_{n}^{\alpha}- \lambda_{n}^{\alpha+1})\tau&=& -2\tau\left(e^{i\alpha\pi}\cos(\frac{\pi}{3}(\alpha-1))-e^{i\beta\pi}\cos(\frac{\pi}{3}(\alpha))\right),\nonumber\\
    &\forall& n\in K'=\{0,1,...,5\};\;\text{and}\;\;\alpha=1.
    \end{eqnarray}
   Based on this choice, from the attainability condition (\ref{attainability2}) we obtain
           \begin{equation} \label{thirdattainabiliy}
           \begin{aligned}
        (\lambda_{n}^{1}- \lambda_{n}^{2})\tau&=3 \tau=2\pi(k_{n}^{1}-k_{n}^{2}),\;\;\;\;\;s_{n}^{1}=s_{n}^{2},\\
       (\lambda_{n}^{1}- \lambda_{n}^{2})\tau&=3 \tau=2\pi(k_{n}^{1}-k_{n}^{2})+\pi,\;\;\;\;\;s_{n}^{1}=-s_{n}^{2}=1,\\
       (\lambda_{n}^{1}- \lambda_{n}^{2})\tau&=3 \tau=2\pi(k_{n}^{1}-k_{n}^{2})-\pi,\;\;\;\;\;s_{n}^{1}=-s_{n}^{2}=-1.
          \end{aligned}
          \end{equation}
  \end{itemize} 
Summarizing, given that there are solutions to Eq. (\ref{attainability2}) for the above three cases we could attain PST. However, our numerical investigations indicate only PST happens when the in and out nodes are in the same channel diametrically opposed. This confirms that the analytical solutions are necessary but not sufficient conditions to have PST which to pointed already in section (\ref{MTF}). In other words, in this example we should only deal with Eqs. (\ref{example1condition1}) in order to look for the equivalence between the analytical solutions with the numerical dynamics of the transmission probability as a function of $\tau$ and $\gamma$. Fig. \ref{N8} $a$ illustrates the transmission probability $p_{\tau}$ versus $\tau$ when $N=8$ and for two different values of $\gamma$, between two diametrically opposed nodes in the same channel as $p_{\tau}([0,1],[4,1])$.  It is evident the possibility of PST for both $\gamma=3$ and $\gamma=5$. The analytical attainability  $p_{\tau}([i,p],[j,q])$ for the condition (\ref{example1condition12}) as $\sqrt{2}\gamma \tau=2\pi(k_{1}^{\alpha}-k_{2}^{\alpha})-\pi$ with $\gamma=3$, $\tau=12.59,73.31,134.03,\cdots$ and $k_{1}^{\alpha}-k_{2}^{\alpha}=8,49,90,\cdots$ shows the consistency  with the numerical outcome $p_{\tau}([0,1],[4,1])=1$. The same agreement between analytical and numerical solutions of  $p_{\tau}([0,1],[4,1])=1$ is also seen for $\gamma=5$ where $\tau=43.985,131.955\cdots$ for $k_{1}^{\alpha}-k_{2}^{\alpha}=49,148,\cdots$. More generally, we display the minimum values $\tau_{min}$ at which the probability $p_{\tau}([0,1],[4,1])=1$ versus $\gamma$ for $N=8$ in Fig. \ref{N8}$b$. Values of $\tau_{min}$ saturating the range of the figure means that the probability never reaches $1$. Then  for $\gamma\rightarrow \infty$, i.e. $L\rightarrow 0$, finite values of $t_{min}$ become more frequent depicted in Fig. \ref{N8} $c$. Thus this can be understood with the fact that in such a limiting case the information is not spread over three channels, but remains confined within only one channel.

\section{open boundary on both $N$ and $\alpha$}\label{opennalpha}

The second condition we consider is the open boundary on both $N$ and $\alpha$ (see Fig. \ref{diffboundaries} $b$). In such a case, the Hamiltonian belongs to the symmetric block-Toeplitz tridiagonal  matrices which can be displayed as 
\begin{equation*}
\hat{H}_{\text{ex}} = 
\begin{pmatrix}
\textbf{C}_{0}^{0} & \textbf{C}_{1}^{0} & 0 & \cdots & 0 \\
\textbf{C}_{0}^{1} & \textbf{C}_{0}^{0} & \textbf{C}_{1}^{0} &   & 0 \\
0 & \textbf{C}_{0}^{1} & \ddots & \ddots & \vdots \\
\vdots  &    & \ddots &    & \textbf{C}_{1}^{0}  \\
0 & 0 & \cdots & \textbf{C}_{0}^{1} & \textbf{C}_{0}^{0}  
\end{pmatrix}
\label{blockTST1}
\end{equation*}
with
\begin{eqnarray}\label{matrixopenNAlpha}
C_{0}^{0}=\left(
      \begin{array}{ccc}
       0 & L&0 \\
       L&0&L\\
        0 & L& 0\\
       \end{array}
     \right);\;\;\;\;\;\;\;\;
            C_{0}^{1}=C_{1}^{0}=\left(  
       \begin{array}{ccc}
       J & 0&0 \\
      0 & J&0\\
       0 & 0&J\\
       \end{array}
     \right).
\end{eqnarray}
The eigenvalues of the above matrix $\hat{H}_{\text{ex}}$ are \cite{ABDERRAMANMARRERO2020125324}
\begin{eqnarray}
\lambda_{n}^{\alpha}=2J\cos((n+1)\pi / N+1)+2L\cos(\alpha\pi / 4),
\end{eqnarray}
where $\alpha=1,2,3 $ and $n=0,1,...,N-1$.

\begin{figure}[H]
 \centerline{\includegraphics[width=0.45\columnwidth]{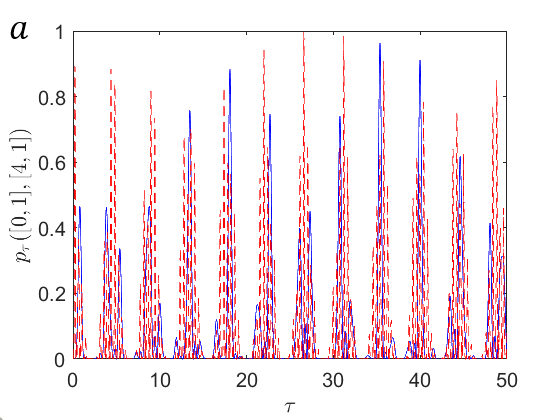}}
  \includegraphics[width=0.45\columnwidth]{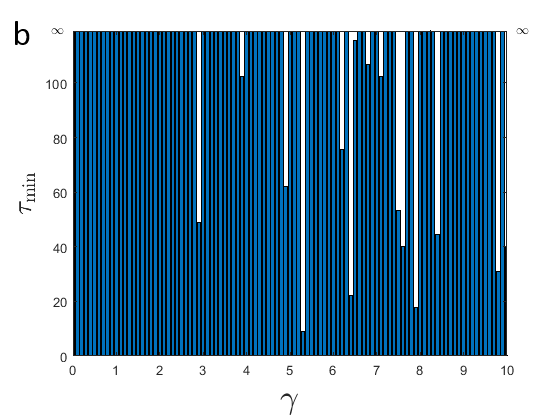}
   \includegraphics[width=0.45\columnwidth]{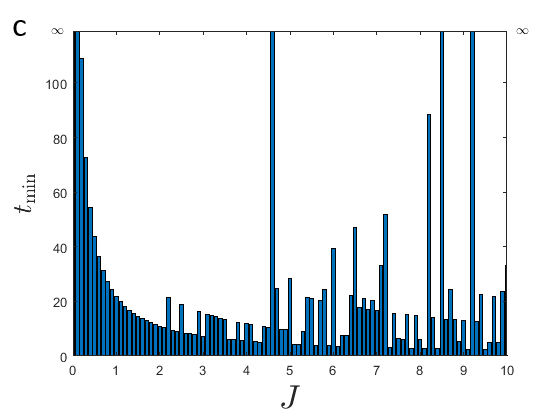}
   \caption{The transmission probability $p_{\tau}([0,1],[4,1])$ of a spin network with $N=5$ nodes and open boundary on both $N$ and $\alpha$; a) $p_{\tau}$ vs $\tau$ for $\gamma=4$ (solid blue line) and $\gamma=15$ (red dashed line), b) the minimum value $\tau_{min}$ versus $\gamma$ at which $p_{\tau}=1$ for the finite values of $\gamma$, and c) the minimum value $t_{min}$ versus $J$ at which $p_{t}=1$ for $\gamma\rightarrow \infty$, i.e. $L\rightarrow 0$.}
\label{N5}
\end{figure}

Given the fact that in this case it is impossible to analytically find the eigenvectors of the Hamiltonian, we are obligated to numerically study the transmission probability $p_{\tau}$ in (\ref{decatt}). Note that only $p_{\tau}([0,1],[4,1])$ and $p_{\tau}([0,1],[4,3])$ enjoy PST and we choose to show $p_{\tau}([0,1],[4,1])$ to have coherency with other boundary conditions. Fig. \ref{N5} shows the probability of the state transmission in the same channel  $p_{\tau}([0,1],[4,1])$ with  the number of sites $N = 5$. We see from Fig. \ref{N5} $a$ the behavior of $p_{\tau}$ versus $\tau$ where PST is observable for $\gamma=15$ when $\tau=26.6,84.4,\cdots$, however there is no PST for $\gamma=4$. Fig. \ref{N5} $b$ displays the smallest time $\tau_{min}$ at which PST is achieved versus the finite values of $\gamma$. As can be seen finite $\tau_{min}$ are quite rare, meaning that PST does not occur for most of the $\gamma$ values. Fig. \ref{N5} $c$ shows that for $L\rightarrow 0$ most of the values of $\tau_{min}$ become finite and actually they generically tend to decrease with $J$.

\section{open boundary on $N$ and closed boundary on $\alpha$}
\label{opennclosea}

\begin{figure}[H]
 \centerline{\includegraphics[width=0.45\columnwidth]{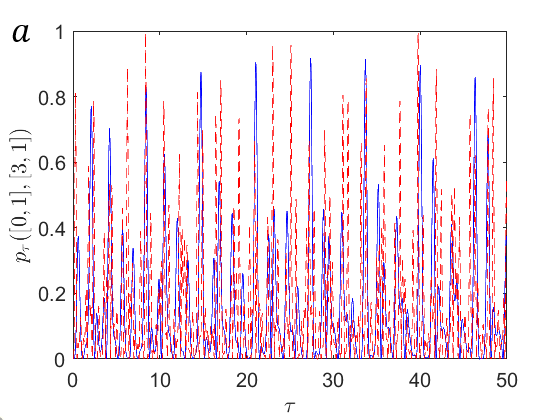}}
  \includegraphics[width=0.45\columnwidth]{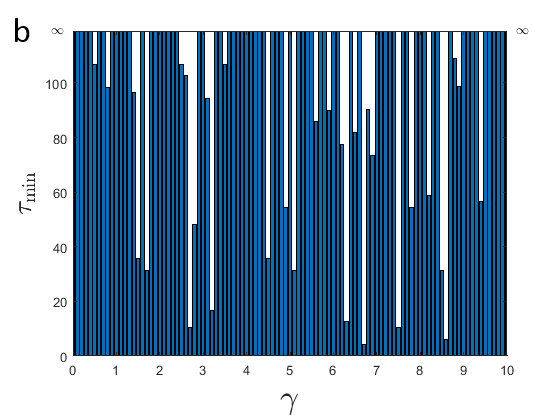}
   \includegraphics[width=0.45\columnwidth]{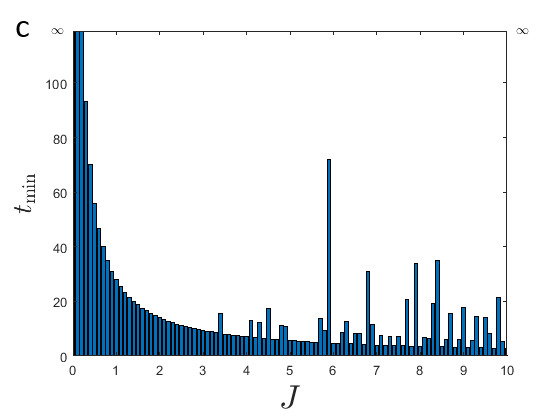}
   \caption{The transmission probability $p_{\tau}([0,1],[3,1])$ of a spin network with $N=4$ nodes and open boundary on $N$ and closed boundary on $\alpha$; a) $p_{\tau}$ vs $\tau$ for $\gamma=4$ (solid blue line) and $\gamma=9.4$ (red dashed line), b) the minimum value $\tau_{min}$ versus $\gamma$ at which $p_{\tau}=1$ for the finite values of $\gamma$, and c) the minimum value $t_{min}$ versus $J$ at which $p_{t}=1$ for $\gamma\rightarrow \infty$, i.e. $L\rightarrow 0$.}
\label{N4}
\end{figure}

The third condition is the open boundary on $N$ and closed boundary on $\alpha$ (see Fig. \ref{diffboundaries} $c$) which again is a block-Toeplitz tridiagonal matrix as in (\ref{blockTST1}); however here 
\begin{eqnarray}\label{matrixH3}
C_{0}^{0}=\left(
      \begin{array}{ccc}
       0 & L&L \\
       L&0&L\\
        L & L& 0\\
       \end{array}
     \right)
\end{eqnarray}
and  $C_{0}^{1}=C_{1}^{0}=J\hat{I}_{3\times 3}$. Then the eigenvalues will be \cite{ABDERRAMANMARRERO2020125324}
\begin{eqnarray}
\lambda_{n}^{\alpha}=2J\cos{((n+1)\pi / N+1)}+2L\cos{(2\pi \alpha / 3)},
\end{eqnarray}
where $\alpha=1,2,3,$ and $n=0,1,...,N-1$. Our investigations shows that PST can be only reachable between the nodes with the longest distance in the same channel for specific $\gamma$. Fig. \ref{N4} shows the transmission probability $p_{\tau}[(0,1),(3,1)]$ with $N=4$. PST can be seen in Fig. \ref{N4} $a$ for $\gamma=9.4$ when $\tau=8.35,39.75,56.5, \cdots$ while it does not occur for $\gamma=4$. Fig. \ref{N4} $b$ shows the least amounts of time to wait for PST versus finite values of $\gamma$. It is observed that PST does not occur for several values of $\gamma$. However Fig. \ref{N4} $c$  is an evidence of well  spreading of information over only one channel due to the finiteness and decreasing behavior of $t_{min}$ versus $J$ when $L\rightarrow 0$, i.e. $\gamma\rightarrow \infty$.  
                   
\section{closed boundary on $N$ and open boundary on $\alpha$}
\label{closenopena}
\begin{figure}[H]
 \includegraphics[width=0.45\columnwidth]{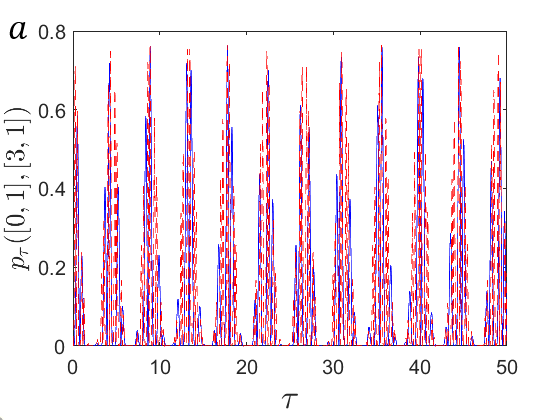}
\includegraphics[width=0.45\columnwidth]{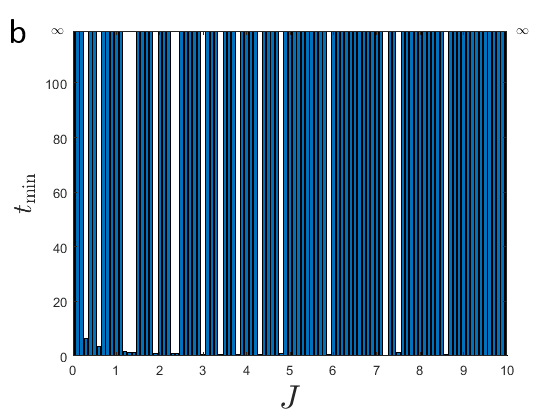}
\caption{The transmission probability $p_{\tau}([0,1],[3,1])$ of a spin network with $N=6$ nodes and closed boundary on $N$ and open boundary on $\alpha$; a) $p_{\tau}$ vs $\tau$ for $\gamma=4$ (solid blue line) and $\gamma=8.25$ (red dashed line), b) the minimum value $t_{min}$ versus $J$ at which $p_{t}=1$ for $\gamma\rightarrow \infty$, i.e. $L\rightarrow 0$.}
\label{N6}
\end{figure}

The last boundary condition to be considered is closed on $N$ and open on $\alpha$. (see fFig. \ref{diffboundaries} $d$). This gives rise to a $3N \times 3N $ block circulant matrix in the form (\ref{BothClose}) with $3 \times 3$ symmetric blocks\\
\begin{eqnarray}\label{matrixHN}
\textbf{C}_{0}^0=\left(
      \begin{array}{ccc}
       0 & L& 0 \\
        L & 0& L \\
        0 & L& 0\\
       \end{array}
     \right);\;\;\;\;\;\textbf{C}_{1}^0=\textbf{C}_{N-1}^0=\left(
      \begin{array}{ccc}
       J & 0& 0 \\
        0 & J& 0 \\
        0 & 0& J\\
       \end{array}
     \right)
\end{eqnarray}
and $\textbf{C}_{2}^0=\textbf{C}_{3}^0=\cdots=\textbf{C}_{N-2}^0=0$. Our numerical investigations show no attainability for finite values of $\gamma$ between any in and out nodes. The highest transmission probability (around 0.8) is achievable between diametrically opposed nodes in the same channel. In Fig. \ref{N6}, we reported $p_{\tau}[(0,1),(3,1)]$ for a network with $N=6$ Fig. \ref{N6} $a$ shows $p_{\tau}$ versus $\tau$ for two different values of $\gamma$ and Fig.\ref{N6} $b$ displays the minimum time to wait for PST versus the coupling constant J. It can be seen that PST is rare even when $\gamma\to\infty$, i.e. $L\rightarrow 0$. Comparing these outcomes with the ones of section (\ref{closenalpha}) we realize a remarkable effect of boundary on $\alpha$ (when it is closed PST is much more frequent).

\section{Conclusions}
\label{Conclusion}

We have studied information transfer fidelity and PST between nodes in a bio-inspired spin network of an $\alpha$-helix structure of a protein using the Davydov Hamiltonian without the phononic environment by assuming the single excitation subspace. We have worked out the quantum state transmission probability for four different cyclic and non-cyclic boundary conditions on both sites and channels of the $\alpha$-helix network. For only one of the boundaries, it was possible to find the analytical eigenvectors of the Hamiltonian and thus investigate the analytical scale times where PST happens. Numerical computations are in compliance with them. We reported numerically the transmission probability and the possibility of PST for other boundaries which are influenced strongly by the coupling terms in the Hamiltonian because PST needs the coupling parameters to provide the right phase matching, admitting the perfect transfer of both amplitude and phase of a quantum state from one node to another. In so doing, the analytical solutions are found to be sufficient but not necessary conditions with respect to the closed boundary on the both sites $N$ and the channels $\alpha$. In such a case, PST happens frequently during the time when the nodes are only diametrically opposed in the same channel; but considering only one or two of the boundaries open leads to the PST over fewer times, i.e. open boundary on both $N$ and $\alpha$, and open boundary on $N$ and closed boundary on $\alpha$. Finally, no attainability is observed for the case of closed boundary on $N$ and open boundary on $\alpha$, nevertheless the highest transmission probability happens for the diametrical opposition of the input and out nodes in the same channel. Futher developments of this model could take into account the effects of local or quasi-local  enviroments \cite{nourmandipour2016entangiling}. Looking ahead not only quantum information transmission, but also quantum information processing could be investigated in a model involving electrons distribution \cite{mancini1999quantum}.

\acknowledgments
This work was partially supported by the European
Union’s Horizon 2020 Research and Innovation Programme under Grant Agreement No.
964203 (FET-Open LINkS project).
RF also acknowledges support from the RESEARCH SUPPORT PLAN 2022 - Call for applications for funding allocation to research projects curiosity driven (F CUR) - Project "Entanglement Protection of Qubits’ Dynamics in a Cavity"– EPQDC and the support by the Italian National Group of Mathematical Physics (GNFM-INdAM).

\section*{Data availability}
All relevant data have been included in the paper.

\section*{Conflict of Interest}
The authors declare that they have no conflict of interest.

  \bibliographystyle{apsrev4-1}
  \bibliography{RefBank}


\end{document}